\documentclass[twocolumn, rnaas]{aastex631}

\usepackage{blindtext}
\usepackage{hyperref}
\hypersetup{
    colorlinks=true,
    linkcolor=blue,
    filecolor=cyan,      
    urlcolor=magenta,
    citecolor=blue,
}

\chardef\us=`\_
\newcommand{\Rs}{\(\mathrm{R}_\odot\)}

\submitjournal{RNAAS}

\shorttitle{New Visualization of Polarized Coronal Images from the 2023 TSE}

\begin{document}

\title{A Chromatic Treatment of Linear Polarization in the Solar Corona at the 2023 Total Solar Eclipse}


\author[0000-0001-8504-2725]{Ritesh Patel}
\affiliation{Southwest Research Institute, Boulder, CO, USA}

\author[0000-0002-0494-2025]{Daniel B. Seaton}
\affiliation{Southwest Research Institute, Boulder, CO, USA}

\author[0000-0001-8702-8273]{Amir Caspi}
\affiliation{Southwest Research Institute, Boulder, CO, USA}

\author[0000-0003-1714-5970]{Sarah A. Kovac}
\affiliation{Southwest Research Institute, Boulder, CO, USA}


\author[0009-0008-4901-0601]{Sarah J. Davis}
\altaffiliation{Now graduated and no longer at this institution}
\affiliation{University of Northern Colorado, Greeley, CO, USA}

\author[0000-0001-6535-3390]{John P. Carini}
\affiliation{Indiana University, Bloomington, IN, USA}

\author[0009-0003-8984-2094]{Charles H. Gardner}
\affiliation{Rice University, Houston, TX, USA}

\author[0000-0002-5504-6773]{Sanjay Gosain}
\affiliation{National Solar Observatory, Boulder, CO, USA}

\author[0000-0002-9714-753X]{Viliam Klein}
\affiliation{Southwest Research Institute, Boulder, CO, USA}

\author[0009-0006-7389-2419]{Shawn A. Laatsch}
\affiliation{University of Maine, Orono, ME, USA}

\author[0000-0002-8043-5682]{Patricia H. Reiff}
\affiliation{Rice University, Houston, TX, USA}

\author[0009-0001-9703-7859]{Nikita Saini}
\affiliation{University of Maine, Orono, ME, USA}

\author[0009-0008-7102-9771]{Rachael Weir}
\altaffiliation{Now graduated and no longer at this institution}
\affiliation{Indiana University, Bloomington, IN, USA}

\author[0000-0002-4425-3443]{Daniel W. Zietlow}
\affiliation{National Center for Atmospheric Research, Boulder, CO, USA}

\author{David F. Elmore}
\affiliation{National Solar Observatory, Boulder, CO, USA}

\author[0009-0006-1260-2297]{Andrei E. Ursache}
\affiliation{Independent Researcher}

\author[0000-0002-7164-2786]{Craig E. DeForest}
\affiliation{Southwest Research Institute, Boulder, CO, USA}

\author[0000-0002-0631-2393]{Matthew J. West}
\affiliation{Southwest Research Institute, Boulder, CO, USA}

\author{Fred Bruenjes}
\affiliation{Daystar Filters LLC, Warrensburg, MO, USA}

\author{Jen Winter}
\affiliation{Daystar Filters LLC, Warrensburg, MO, USA}

\author{the Citizen CATE 2024 Team}



\begin{abstract}
The broadband solar K-corona is linearly polarized due to Thomson scattering. Various strategies have been used to represent coronal polarization. Here, we present a new way to visualize the polarized corona, using observations from the 2023~April~20 total solar eclipse in Australia in support of the Citizen CATE 2024 project. We convert observations in the common four-polarizer orthogonal basis (0\degr, 45\degr, 90\degr, \& 135\degr) to $-60\degr$, 0\degr, and +60\degr\ (\emph{MZP}) polarization, which is homologous to \emph{R, G, B} color channels. The unique image generated provides some sense of how humans might visualize polarization if we could perceive it in the same way we perceive color. 
\end{abstract}

\keywords{Sun: corona --- Sun: Solar K corona --- Solar Eclipses  --- Astronomical techniques: Polarimetry}

\section{Introduction} 
\label{sec:intro}
The linear polarization of the solar corona is often used both to remove unpolarized background brightness from broadband visible-light coronal images and to disentangle 3D structure \citep{Deforest2017}. Close to the Sun, the K-corona is strongly polarized and the F-corona is essentially unpolarized. For eclipse studies, polarimetric observations can be used to completely isolate the structure of the K-corona, enabling high-contrast, high-resolution imaging of the inner and middle corona \citep{West2023} -- a region that is otherwise very difficult to observe.

The K-corona primarily results from Thomson-scattered light, and therefore the polarization angle is determined by the geometric relationship between the Sun, the observer, and the scattering point. Neglecting (scientifically important) deviations due to the 3D complexity of the corona, the polarization angle of K-coronal light near the Sun is tangent to the solar disk in the plane of the sky. This causes a characteristic pattern in the polarization of the corona as a function of position angle around the Sun, which -- in addition to its scientific applications -- can be used to validate the performance of polarimetric observations. \citep[See][for a thorough discussion.]{DeForest2022ApJ}

The Citizen CATE 2024 project \citep{Kovac2022} will leverage coronal polarization to characterize structure and dynamic processes in the visible-light corona. CATE 2024 will observe the Sun from 35 stations, operated by community participants (``citizen scientists''), along the 2024~April~8 total solar eclipse (TSE) track. As an initial test of the CATE 2024 equipment and procedures, our team traveled to the 2023~April~20 TSE in Australia to observe the polarized corona. Here, we provide an initial report of our observations and present a novel approach to visualizing coronal polarization that serves as a validation of our measurements.

\section{Observations}
\label{sec:obs}

We observed the TSE from Exmouth, Western Australia (21\degr 56\arcmin 43.97\arcsec\,S 114\degr 07\arcmin 56.64\arcsec\,E), in the North West Cape, which was the only Australian region in the path of totality. Totality at our location began at 03:29:47\,UT and lasted for 58.3\,s. The results shown in this paper were obtained just before third contact at the end of totality.

The CATE 2024 observing setup includes a polarization-sensitive camera, a FLIR Blackfly BFS-PGE-123S6P-C; a doublet ED refractor telescope with 80\,mm aperture and focal ratio of \textit{f}/6.25, a Long Perng S500G-A with a DSUV2 UV/IR cut filter, both procured from Daystar Filters; and an iOptron GEM28 German Equatorial tracking mount and tripod. The FLIR camera uses a Sony IMX253MZR sensor, which provides images with 4096$\times$3000 pixels, where each 2$\times$2 macropixel uses on-chip filters to observe polarization angles of 0\degr, 45\degr, 90\degr, and 135\degr. The optics coupled with the camera provide a field of view (FOV) of 1.63\degr$\times$1.19\degr (or $\pm$3.07$\times\pm$2.25\,\Rs\ at the time of the eclipse). We used a set of eight logarithmically-spaced exposures (0.123--399.995\,ms) to cover the three orders of magnitude of dynamic range of the corona in our FOV. 

\section{Analysis and Results}
\label{sec:results}

\begin{figure*}
    \centering
    \includegraphics[width=1\linewidth]{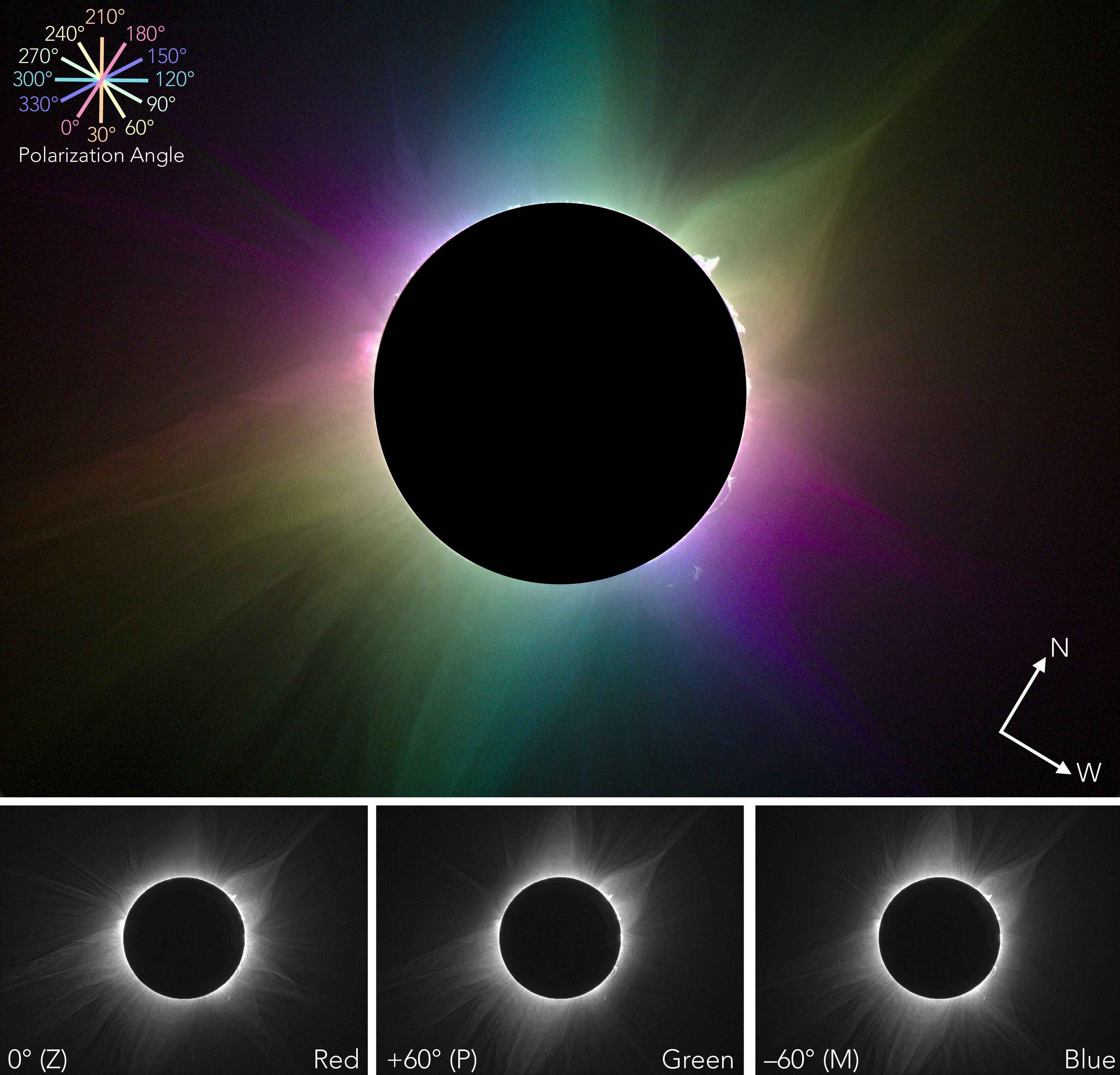}
    \caption{Linearly polarized broadband visible-light K-corona from the 2023~April~20 total solar eclipse observed by Citizen CATE 2024 in Exmouth, Western Australia. The central disk is an overlay and is slightly oversized (105\%) to exclude limb-edge artifacts. The colors shown here are a linear combination of red, green, and blue, analogous to \textit{MZP} polarization states. Solar north and west are indicated at bottom right; color-angle correspondence is indicated in the upper left. The small sub-panels show the individual virtual Z, P, and M images, respectively, used to generate the RGB polarization image.}
    \label{fig:final_img}
\end{figure*}

To produce the final image (Figure~\ref{fig:final_img}), we separate raw frames into four 2048$\times$1500-pixel arrays corresponding to the individual on-chip polarizer angles. These images are reduced and multiple exposures are combined to create a single high-dynamic-range (HDR) image of the solar corona at each polarizer angle. The images at all four angles are then converted to triplets using the \textsf{solpolpy} solar polarization resolver Python package \citep{marcus2023, Walbridge2022AGU}, based on the conversions described in \citet{DeForest2022ApJ}. The triplet images use basis angles of $-60\degr$, 0\degr, and +60\degr, which we refer to as \textit{MZP} (for Minus, Zero, and Plus). 

We used the \textsf{Multi-scale Gaussian Normalization} package \citep[MGN;][]{Morgan2014SoPh} to process the HDR images to enhance the fine-scale coronal structures. Separately, to manage the dynamic range across large scales in the image, we applied the SWAP azimuthally-varying radial filter \citep{Seaton2023SoPh} to each HDR image. We combine these SWAP-filtered images of large-scale coronal structures with the MGN-processed high-frequency-enhanced images using an ad-hoc intensity scaling to obtain natural-looking eclipse images with enhanced details at multiple spatial scales.  

We merged the three individual processed \textit{MZP} images into a single color image by respectively mapping each individual frame into one of the three RGB channels. We increased the saturation by 50\% to enhance the color contrast of the final image shown in Figure~\ref{fig:final_img}. 

We selected our \textit{MZP} triplet such that \textit{Z} (0\degr) maps to the red channel and is aligned with solar north, thus red appears at the east limb where the tangent direction is parallel to the solar polar axis. As position angle increases (counter-clockwise) the colors progress from red to green to blue from solar north to south. This cycle is repeated on the west limb of the Sun, due to the 180\degr\ ambiguity in polarization angle. Intermediate colors represent the amount of departure of a particular polarization direction from one of the primary \textit{MZP} components. The prominences appear white, indicating roughly equal brightness in each of the polarization channels due to their largely unpolarized emission.

\section{Discussion}
\label{sec:conclusion}

Unlike conventional methods that assign single colors to individual polarizer angles and blend the respective images, our method incorporates the linear combination of polarization represented by colors according to color theory \citep{Smith1931TrOS}. Our results demonstrate the utility of the RGB analog of polarization to vividly visualize structure, brightness, and polarization in a single, high-contrast image. They also serve to validate the excellent performance of our CATE 2024 equipment using real eclipse data. 

If the human visual system were sensitive to linear polarization, as we are to colors, then we might perceive the solar corona similarly to Figure~\ref{fig:final_img}. Additional processing to reveal small variation in polarization angle around the Sun will help to unravel the underlying 3D structure and density of the corona. An in-depth analysis of the entire Australian TSE data set will be presented in a subsequent study.

\begin{acknowledgements}
Funding for Citizen CATE 2024 was provided by grants from the National Science Foundation (award numbers 2231658, 2308305, \& 2308306) and NASA (grant numbers 80NSSC21K0798 \& 80NSSC23K0946). We thank Teledyne FLIR LLC and Daystar Filters LLC for their contributions. Data from the 2023 TSE in Australia is available on request at \url{https://eclipse.boulder.swri.edu}.
\end{acknowledgements}

\bibliography{reference}{}
\bibliographystyle{aasjournal}



\end{document}